\documentclass[letterpaper, 10 pt, conference]{./ieeeconf}  %

\IEEEoverridecommandlockouts                              %

\usepackage{graphicx}
\usepackage{amsmath} %
\usepackage{amssymb}  %
\usepackage[nohyperlinks,printonlyused,nolist]{acronym}
\usepackage{booktabs}
\usepackage{balance}

\usepackage{siunitx}
\usepackage{tikz}
\usetikzlibrary{calc, positioning}
\newif\ifpreprint

\def\conferenceyear{2019}                                       %
\def\conferencetitle{2019 IEEE International Symposium on Safety, Security, and Rescue Robotics}      %
\def\conferencetitleabrv{SSRR}                            %

\def\conferencenotice{
	Accepted for presentation in: \conferencetitle~(\conferencetitleabrv).}
\def\copyrightnotice{
	\copyright~\conferenceyear~IEEE. Personal use of this material is permitted. Permission from IEEE must be obtained for all other uses, including reprinting/republishing this material for advertising or promotional purposes, collecting new collected works for resale or redistribution to servers or lists, or reuse of any copyrighted component of this work in other works.}

\preprinttrue

\title{\LARGE \bf
Robust Cellular Communications for Unmanned Aerial Vehicles in Maritime Search and Rescue
}

\author{Philipp Gorczak$^1$, Caner Bektas$^1$, Fabian Kurtz$^1$, Thomas L{\"u}bcke$^2$ and Christian Wietfeld$^1$%
\thanks{$^{1}$Philipp Gorczak, Caner Bektas, Fabian Kurtz and Christian Wietfeld are with the Communication Networks Institute, TU Dortmund University, 44227 Dortmund, Germany
{\tt\small \{Philipp.Gorczak, Caner.Bektas, Fabian.Kurtz, Christian.Wietfeld\}@tu-dortmund.de}}%
\thanks{$^{2}$Thomas L{\"u}bcke is with German Maritime Search
and Rescue Service (DGzRS), Werderstrasse 2, 28199 Bremen, Germany
{\tt\small luebcke@seenotretter.de}}%
}

\begin{document}

\begin{acronym}
    \acro{BSR}{buffer status report}
    \acro{FDD}{frequency division duplex}
    \acro{LTE}{Long Term Evolution}
    \acro{MAC}{medium access control}
    \acro{MCS}{modulation and coding scheme}
    \acro{OFDMA}{orthogonal frequency-division multiple access}
    \acro{PRB}{physical resource block}
    \acro{PTP}{Precision Time Protocol}
    \acro{QoS}{quality of service}
    \acro{SAR}{search and rescue}
    \acro{SDR}{software-defined radio}
    \acro{SNR}{signal-to-noise ratio}
    \acro{SPS}{semi-persistent scheduling}
    \acro{TBS}{transport block size}
    \acro{TDD}{time division duplex}
    \acro{UAS}{unmanned aerial system}
    \acro{UAV}{unmanned aerial vehicle}
\end{acronym}

\maketitle
\thispagestyle{empty}
\pagestyle{empty}

\ifpreprint
    \begin{tikzpicture}[remember picture, overlay]
    	\node[below=5mm of current page.north, text width=20cm,font=\sffamily\footnotesize,align=center] {\conferencenotice};
    	\node[above=5mm of current page.south, text width=15cm,font=\sffamily\footnotesize] {\copyrightnotice};
    \end{tikzpicture}%
\fi

\begin{abstract}

Unmanned aerial vehicles (UAV) are a promising technology for fast, large scale survey operations such as maritime \ac{SAR}.
However, providing reliable communications over long distances remains a challenge.
Cellular technology such as \ac{LTE} is designed to perform well in the presence of fast-moving clients and highly dynamic channel conditions.
Yet, its generic medium access mechanisms make it hard to support low-latency uplink traffic, which is needed for real-time command and control and similar applications.
In this paper, we introduce an unmanned aerial system for maritime \ac{SAR}, supported via an \ac{LTE} data link.
To mitigate drawbacks in uplink channel access, we develop a resource-guaranteed scheme based on persistent scheduling, using an open-source \ac{LTE} stack.
The approach is evaluated with a laboratory setup using \acl{SDR} modules.
Measurements are taken on the application layer, using real-world telemetry data generated by an autopilot system in the presence of high bandwidth background traffic.
Evaluations demonstrate that our system fulfills UAV requirements in terms of latency and reliability.
Latency is further reduced to \SI{6}{\milli\second} on average and a 99.9th percentile of \SI{10}{\milli\second} by application-aware overprovisioning for mission-critical traffic and users.

\end{abstract}
\acresetall
\section{INTRODUCTION}
\label{sec:intro}

Unmanned aerial vehicles \acused{UAV}(\acp{UAV}) can survey large areas within short timespans, making them ideal platforms for maritime \ac{SAR} operations.
However, real-time monitoring and remote operation of these unmanned systems require robust and reliable communications.
In this context, cellular technologies have emerged as a promising solution for connecting robotic systems in the field due to their capabilities in handling mobility and channel variability.
The current generation, \ac{LTE}, offers a flexible physical layer as well as high capacity and efficiency due to centrally managed channel access. %
Compared to technologies operating in shared, unlicensed frequency bands, centralized scheduling enables \ac{QoS} management, e.g. prioritization of specific traffic flows or resource guarantees.
In addition, the ubiquity of cellular devices and the wide availability of highly optimized hardware modules set \ac{LTE} apart from proprietary licensed band technologies.
Cellular communication is designed to support a variety of traffic types with different requirements (e.g., data and voice). Therefore, its use could obviate the common architectural choice in \acp{UAV} of using independent links for mission-critical and payload communications, each with dedicated hardware.
The adoption of \ac{LTE} for robotic systems control, however, is precluded by its comparatively high latency, resulting from a system design focused on resource efficiency.
The \ac{LTE} standard does support a set of \ac{QoS} classes that can be used to identify traffic with specific latency or data rate requirements~\cite{3gpp-ts23.203r14}.
Yet, none of these profiles satisfy the demands imposed by command and control links, which require one-way delays of \SI{50}{\milli\second}, and a high reliability of \SI{99.9}{\percent} \cite{3gpp-ts36.777r15}\cite{7463007}. %

\begin{figure}[b]
\vspace{-4mm}
    \centering
    \includegraphics[width=1\columnwidth]{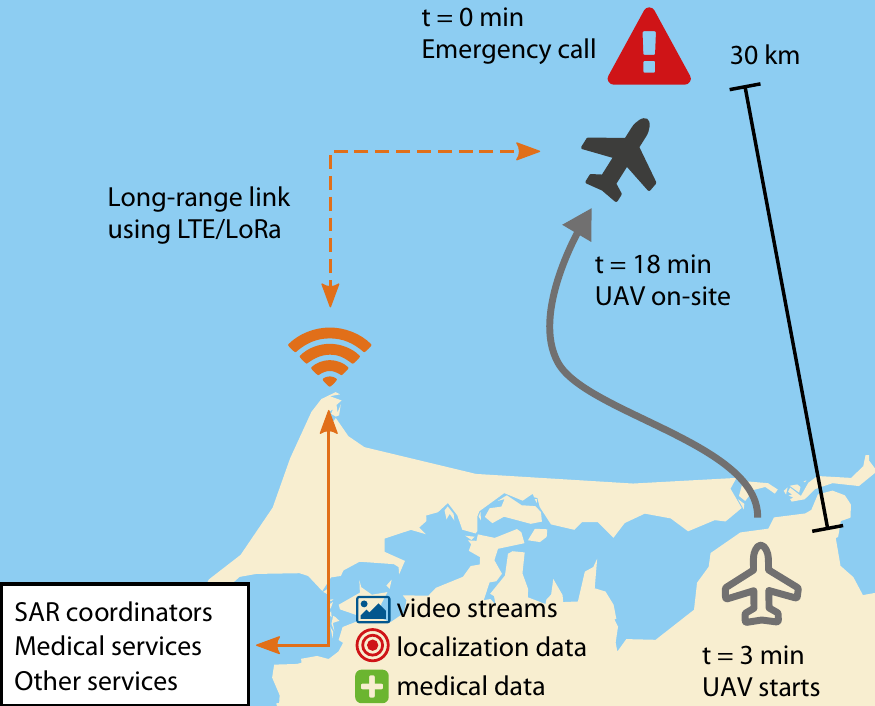}
    \caption{Illustration of the proposed system. The early start and high travel speed of the UAV reduces the on-site arrival time significantly.}
    \label{fig:scenario}
\end{figure}

Appropriate scheduling strategies at the \ac{MAC} layer are required to enable the provisioning of any latency guarantees for mission-critical communications.
General-purpose schedulers, as discussed in related works, predominantly optimize resource allocation based on knowledge of per-packet sizes and arrival times.
However, signaling of this information is not possible in the \ac{LTE} control plane.
In addition, \ac{LTE}'s mechanism for signaling data volume is in itself a grant-based uplink process, further increasing latency if schedulers require this information before allocating resources.
By incorporating knowledge of users' mean throughput requirements, this work proposes and investigates a persistent, periodic resource allocation approach, providing hard service guarantees for robust control of \acp{UAV}.
Motivated by the use of \acp{UAV} in maritime \acl{SAR}, we investigate how the delay and reliability of \ac{LTE}-based control links can be improved by allocating dedicated physical resources.
Trade-offs between system capacity and latency are analyzed by comparing different allocation schemes, while robustness is evaluated by introducing competing data traffic.

The remainder of this work is organized as follows.
First, an overview of our concept for a maritime unmanned aerial search and rescue system is given.
An analysis of related work in the areas of \ac{UAV}-aided \acl{SAR} and \ac{UAV} communications follows.
We then describe our proposed \ac{LTE} uplink and scheduling strategies for low latency communications, leading to experimental evaluation of the trade-off between latency, reliability, and capacity.
Although proposed scheduling strategies are static, a perspective on adaptive allocation is presented.
Finally, a discussion of results and an outlook on future work conclude the paper.
\section{MOTIVATION AND SYSTEM OVERVIEW}
\label{sec:system}
When an emergency call arrives at a maritime rescue station, a fast response is key. %
Current water-based, manned \ac{SAR} strategies are highly developed.
However, boats are comparably slow and can not enter some coastal areas due to vegetation or shallow water.
Aerial search is carried out using helicopters, which involves high costs and is impossible during high winds or strong precipitation.

The LARUS research project aims to provide mission support through the deployment of a \ac{UAV}, as illustrated in Fig.~\ref{fig:scenario}.
Traveling at an airspeed of up to \SI{56}{\meter/\second}, LARUS-aircraft reach the operating site considerably faster than boats, while also being able to operate in strong winds that are unsafe for helicopters.
Coastal and territorial waters are covered with a planned operational radius of \SI{30}{\kilo\meter}.

\subsection{Platform and Payloads}
The LARUS-platform is a fixed-wing \ac{UAV} with a combustion engine and \SI{25}{kg} take-off weight.
The aircraft structure is modified for deployment in harsh weather and maritime environments.
Air pressure sensors and high-performance actuators can actively compensate wind gusts during operation in adverse atmospheric conditions~\cite{islam2018design}.
Since search and rescue operations will be supported from uncontrolled airspace and beyond line of sight, the system is equipped with secondary radar transponders.
In addition, a monopulse radar suite is used to avoid collision with passive obstacles~\cite{8448240}.
Detection of persons and objects in the water is provided by a gimbal-mounted suite of optical sensors including visible light, thermal vision, and actively illuminated near-infrared.
Images are pre-classified on-board and forwarded if they contain detections.
The system is equipped with a standalone \ac{LTE} base station~\cite{8417761} that can provide additional communication resources at the site of operation.
This component also enables the localization of users based on radio signal strength~\cite{8644443}.

\subsection{Communications}
Communication with the ground station is provided through two different technologies: a redundant, low power wide area (LPWA) link in unlicensed bands, and LTE, using a dedicated base station.
We choose the proprietary \textit{LoRa} physical layer for the LPWA link, motivated by its high robustness and sensitivity while still complying with unlicensed band power restrictions.
The provided data rate is sufficient for low-rate telemetry and control (\SI{5}{kbps}) without delay requirements.
The dedicated LTE link can be used for critical and high volume traffic with higher power usage and access to licensed spectrum. The choice of LTE is motivated by the wide availability of high-quality modems and the potential of using public networks in the future.
The LTE link is required to transport heterogeneous traffic with varying requirements.
While high-resolution images of detected objects are not delay sensitive, other data such as input and feedback for camera teleoperation as well as relayed voice data need highly reliable low latency transport.
Furthermore, when the link quality degrades e.g., at the edge of the cellular coverage, critical traffic should be prioritized over non-critical traffic. These considerations require a specialized scheduler at the base station, which we propose in section~\ref{sec:scheduling}.

\section{RELATED WORK}
\label{sec:relwork}
Support of search and rescue through robots is both an active field of research and increasingly applied in practice~\cite{murphy2016disaster}.
Recent deployments of \acp{UAV} for assistance at disaster sites include flooding events~\cite{7017670,7784277} as well as during and after Hurricane Harvey~\cite{8468647}.
While these land-based operations mainly rely on rotorcraft, existing works aiming at maritime deployment mainly utilize fixed-wing platforms.
Furthermore, Zolich et al. describe a system architecture and present experimental results for long-range \ac{UAV} operations at sea~\cite{7996481}.
These works utilize WiFi or proprietary communications solutions.
A project closely aligned with the LARUS approach is introduced in~\cite{7271427}, where a fixed-wing \ac{UAV} with an optical sensor suite supports maritime \ac{SAR} tasks.
The authors identify limitations in communication range and data rate, restricting ground-based monitoring, as targets for future research.

\begin{figure*}[tb]
    \centering
    \includegraphics[width=0.95\textwidth]{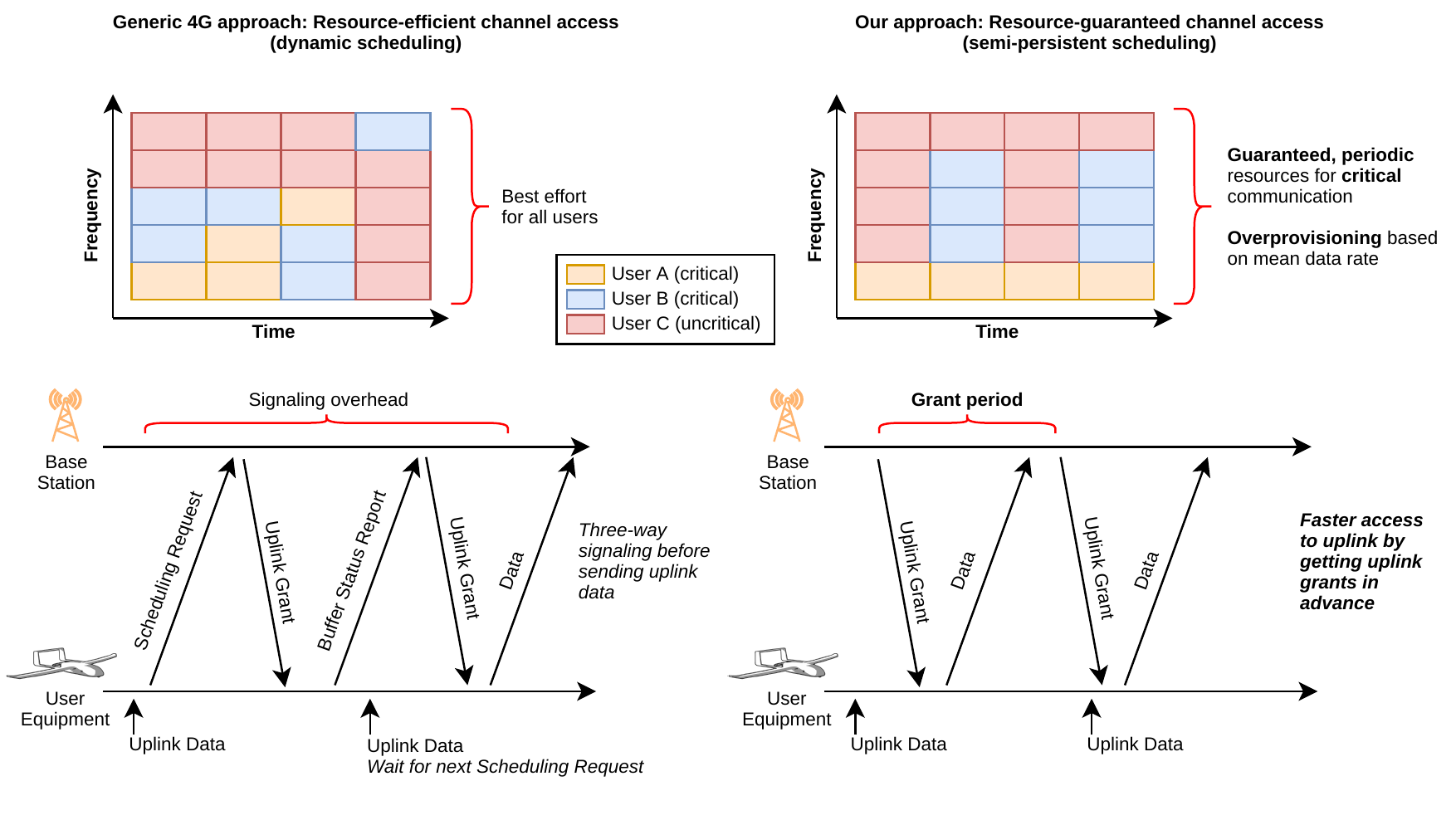}
    \vspace{-6mm}
    \caption{Comparison of LTE resource allocation using default signaling and scheduling to our proposed application-aware persistent allocation.}
    \label{fig:scheduling_overview}
    \vspace{-4mm}
\end{figure*}

While the works mentioned above predominantly focus on point-to-point links, integration of \acp{UAV} into multi-user networks has emerged as another topic of interest~\cite{7463007}.
Cellular technology has been proposed due to the high mobility of \acp{UAV} and the need to cope with dynamic channel conditions.
Unique challenges faced by cellular \ac{UAV}-clients are reported in~\cite{8470897}, including the uplink-heavy asymmetry of traffic, and the need for reliable low latency messaging.

Since current networks are optimized for terrestrial users, network design aspects of cellular-connected \acp{UAV} have been studied in recent works. Models for network analysis covering terrestrial and aerial users are developed in \cite{8692749}. Methods for robust integration of \acp{UAV} into terrestrial LTE networks are investigated in \cite{8301389}. Finally, \cite{8630650} proposes a novel network architecture that increases coverage probability for aerial users. While coverage maximization is an essential prerequisite for reliable cellular connectivity of aerial users, \ac{QoS} guarantees also require appropriate scheduling algorithms. Deadline-aware general-purpose schedulers are evaluated in \cite{6705373}. These algorithms require the age of pending data as input, which can not be signaled through standard control messages. The authors propose to modify the \ac{BSR} message to include the age of the oldest pending packet, thereby making the scheduler incompatible with existing networks. An alternative approach incorporating historical data of standard information is given by \cite{MAIA201675}. While the aforementioned work provides soft guarantees as well as fairness using an optimization algorithm, the approach proposed in this paper investigates a low complexity scheduling method designed for hard guarantees.
In~\cite{7733543}, the authors propose \ac{SPS} for low latency factory automation and evaluate it via simulations for a basic periodic traffic pattern.
On the other hand, this work employs realistic traffic, i.e., generated by an auto-pilot, evaluating the performance of our implementation in a fully functional \ac{LTE} stack.

\section{LOW LATENCY UPLINK SCHEDULING} %
\label{sec:scheduling}
This section gives an overview of scheduling and related signaling procedures in \ac{LTE} and describes the proposed resource allocation scheme.
The remainder of this paper assumes that the system operates in a \ac{FDD} configuration.
However, the proposed approach is transferable to \ac{TDD} operation with minor adjustments reflecting the temporal frame structure and an appropriate uplink-downlink-ratio.

\subsection{LTE Resource Grid and Grant Mechanism}
The \ac{LTE} physical layer is based on \ac{OFDMA}, allowing multiple users to access different spectrum parts concurrently.
Allocations in the frequency domain are valid for \SI{1}{ms} (a \textit{subframe}) and span multiples of 180~kHz.
This time-frequency-division scheme forms a grid of \textit{\acp{PRB}} which is visualized in the upper part of Fig.~\ref{fig:scheduling_overview}.
Cell bandwidth determines the total number of available resource blocks.
Commonly, bandwidths of 5, 10, 15 and \SI{20}{MHz} are used, respectively supporting 25, 50, 75 and 100~\acp{PRB} per subframe.
On the \ac{MAC} layer, the \ac{LTE} base station allocates \textit{grants} within the resource grid, signaling to users which specific \acp{PRB} may be used to transmit data. %
The volume of data transferable within a single grant is known as \textit{\ac{TBS}}.
It is determined by the number of contiguous resource blocks and the employed \ac{MCS}.
The latter are indexed from 0 to 28 and allow adjusting spectral efficiency (data per bandwidth) based on \ac{SNR}.
The mapping from number of resource blocks~$n$ and \ac{MCS}~$m$ to data volume in bits~$v$,
\begin{equation} \label{eq:tbs}
\mathrm{tbs}(n, m) = v
\end{equation}
is defined by the \ac{LTE} standard \cite{3gpp-ts36.213r14}.
It is worth noting that the spectral efficiency of a grant increases up to a certain size due to constant overhead, after which it approaches a value determined by the \ac{MCS}. Block size can furthermore vary by more than an order of magnitude depending on \ac{MCS}.

\subsection{Uplink Signaling and Transfer}
\ac{LTE}'s uplink is based on three procedures: scheduling requests, buffer status reports, and user plane data transfers.
Scheduling requests are periodic opportunities at which a user signals whether there is pending uplink data via a single bit.
Upon reception of a positive request, the base station schedules, and triggers an uplink control plane message indicating the volume of data to be sent, called buffer status report.
After receiving the report, the base station can allocate appropriate grants.
By communicating the need to upload data and the amount of pending data in separate steps before sending the actual data, the LTE uplink architecture enables efficient resource utilization at the expense of increased latency.
As a consequence, the uplink is affected by three different sources of delay.
When an uplink packet arrives at the user, it may have to wait until the next scheduling request period (usually \SI{10}{ms}) to signal its need for uplink resources.
Grants, as required by buffer status reports and the user plane data, are pre-dated by \SI{4}{ms}, further increasing latency.
Implementation-dependent processing time may cause additional delays at each point in the procedure.
The lower section of Fig.~\ref{fig:scheduling_overview} illustrates the uplink scheduling and transfer mechanism.
Measurements in live networks show a total median delay of \SI{30}{ms} with a lower bound of \SI{\sim10}{ms}~\cite{6260469}.

\subsection{Proposed Resource Allocation Scheme} %
This work proposes a static resource allocation scheme for delay-sensitive data flows.
The scheduling request and buffer status report procedures are avoided by granting uplink resources periodically, without requiring prior signaling.
This approach is tailored to critical use cases, featuring traffic with known average data rates, such as voice data or control and telemetry data of unmanned vehicles.

Resources are overprovisioned by a set factor to accommodate traffic bursts above the average data rate. While reducing the overall capacity of the system, this allocation of additional resources is required to avoid excessive delays during traffic fluctuations and to decrease the variation of packet delays (jitter).

Grants are calculated based on the mean data rate~$r$, a period~$p$, and the provisioning factor $\alpha \geq 1$.
Every~$p$ subframes, users receive a grant of size~$n$, with

\begin{equation} \label{eq:grant}
n = \min\{n' \mid \mathrm{tbs}(n', m) \geq \frac{r p \alpha}{1000} \}.
\end{equation}

The users supported by a cell with $N$~\acp{PRB} is given by

\begin{equation} \label{eq:grant_scalability}
\lfloor\frac{N}{n}\rfloor p.
\end{equation}

Note that $N \geq n$ is required for an allocation scheme to be feasible and to support at least one user.
A trade-off between capacity and delay is given by the period~$p$.
Allocating larger grants with longer periods increases efficiency due to decreasing overhead.
At the same time, delays increase due to messages waiting up to one period until transmission.

After distributing periodically scheduled resources, the remaining grants are allocated based on buffer status information. Thus high priority and regular users can co-exist while keeping prioritization and service guarantees.

The scheduling scheme described above is a low complexity approach of guaranteeing resources by utilizing knowledge about the supported application. It has two free parameters: the period and the provisioning factor. In the following section, we investigate how these parameters affect the delay and reliability of realistic telemetry traffic, as well as the efficiency of the resource allocation.
\section{EXPERIMENTAL RESULTS}
\label{sec:results}
\begin{figure}[t]
    \centering
    \vspace{2mm}
    \includegraphics[width=1\columnwidth,trim={0 4mm 0 0}]{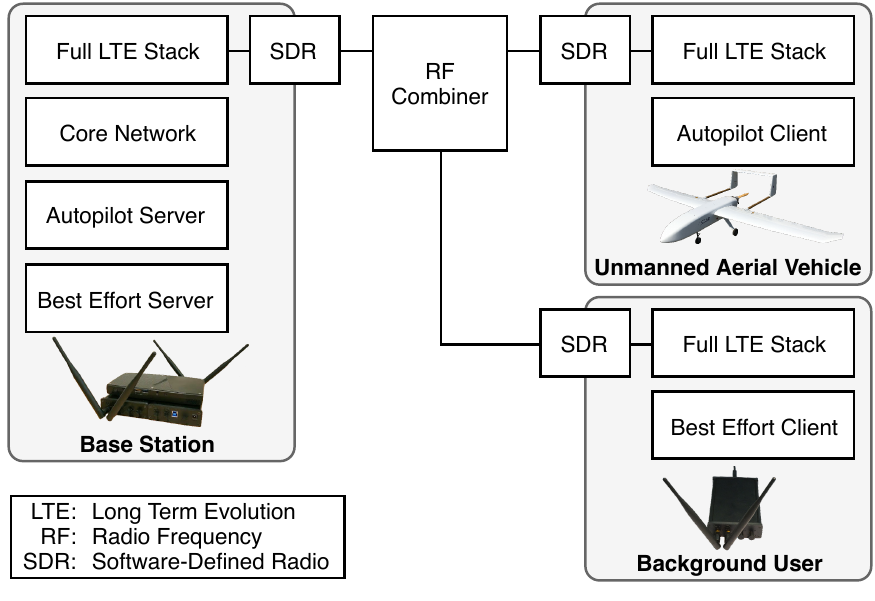}
    \caption{Laboratory setup used for experiments. Delays are measured between Autopilot applications, using local clocks synchronized via Ethernet.}
    \label{fig:setup}
    \vspace{-3mm}
\end{figure}

To evaluate our proposed approach, the scheduling method introduced above is implemented in an automated setup utilizing virtualized components and \ac{SDR} frontends~\cite{8417761}, as depicted in Fig.~\ref{fig:setup}. %
The utilized software stack is fully open source (srsLTE~\cite{Gomez-Miguelez:2016:SOP:2980159.2980163} and NextEPC).
The base station receives scheduling parameters for each user, identifying them as persistent or best-effort clients.
In this implementation, each grant is signaled separately via downlink control information.
\ac{LTE}'s semi-persistent scheduling mechanism can be used to reduce control plane traffic.

Each experiment consists of a \ac{UAV} and background user, connecting to the \ac{LTE} base station and transmitting data.
The \ac{UAV} receives periodic uplink grants as described in the previous section.
The background user generates best-effort traffic, putting the network under full load by requesting more resources than available.
This user is scheduled using buffer information obtained by the standard signaling mechanism.
We use a constant \ac{MCS}~7 to restrict the parameter space and allow for a predictable data rate available to the secondary user.
This choice of \ac{MCS} requires a relatively low \ac{SNR} of \SI{5}{\decibel} on the physical channel~\cite{kawser2012downlink}.

End-to-end delays are measured on the application layer. We use a telemetry data stream originating from an open-source autopilot system (Paparazzi).
The system sends 27 different message types in fixed intervals ranging from \SIrange{0.1}{11.1}{\second} with sizes of \SIrange{10}{40}{Bytes}.
The resulting data stream of 47~messages per second requires a mean data rate of \SI{19.5}{kbps}.
UAV and base station host clocks are synchronized via \ac{PTP} to enable accurate delay measurements.

Grant periods from \SIrange{1}{50}{\milli\second} are evaluated in \SI{3}{\milli\second} steps, with provisioning factors of \numlist{1; 1.5; 2}.
Every test runs for \SI{15}{\minute}, yielding 40,000 data points each.

\begin{figure}[t]
    \centering
    \vspace{2mm}
    \includegraphics{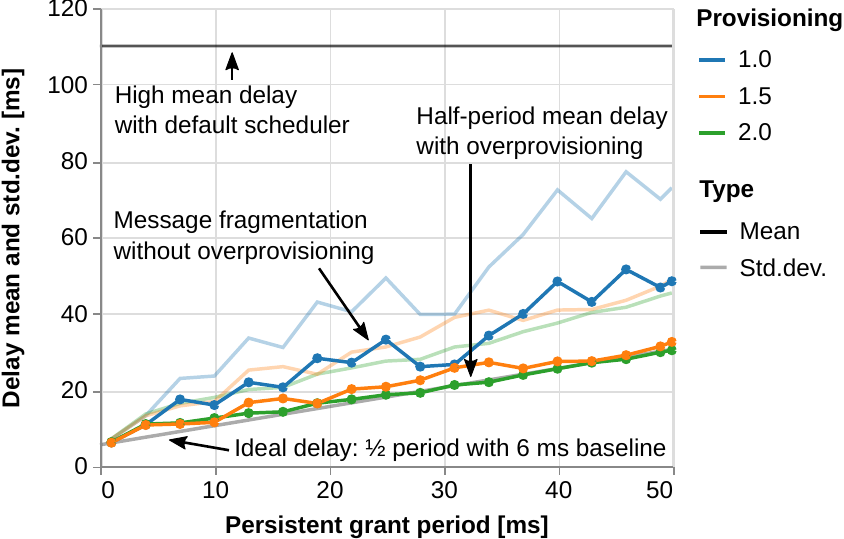}
    \caption{Delay statistics under full network load using our persistent scheduling approach. Mean and upper standard deviation intervals of end-to-end measurements are shown. Latency with default scheduling was measured in our setup but also observed in a field study in~\cite{6857113}.}
    \label{fig:delay_stats}
    \vspace{-4mm}
\end{figure}

\subsection{Latency Statistics and Reliability}
The mean and standard deviation intervals of observed end-to-end delays are shown in Fig.~\ref{fig:delay_stats}.
The best-case mean delay is at \SI{6}{\milli\second}, when resources are granted in every subframe, leaving only processing times and the \SI{4}{\milli\second} delay inherent to \ac{LTE} uplink grants.
For longer grant periods, both mean and variance increase.
It is worth noting that overprovisioning is required to achieve an ideal average delay of half a period. %
When no extra resources are allocated, messages are delayed by about one period on average.
Note that on the lower grant period range, delays deviate more from the ideal model. This effect is due to an increased likelihood of payload fragmentation across multiple small grants. Upwards of the mean inter-arrival time of \SI{21}{\milli\second}, the impact of this inefficiency starts to decrease.

\begin{figure}[t]
    \centering
    \vspace{2mm}
    \includegraphics{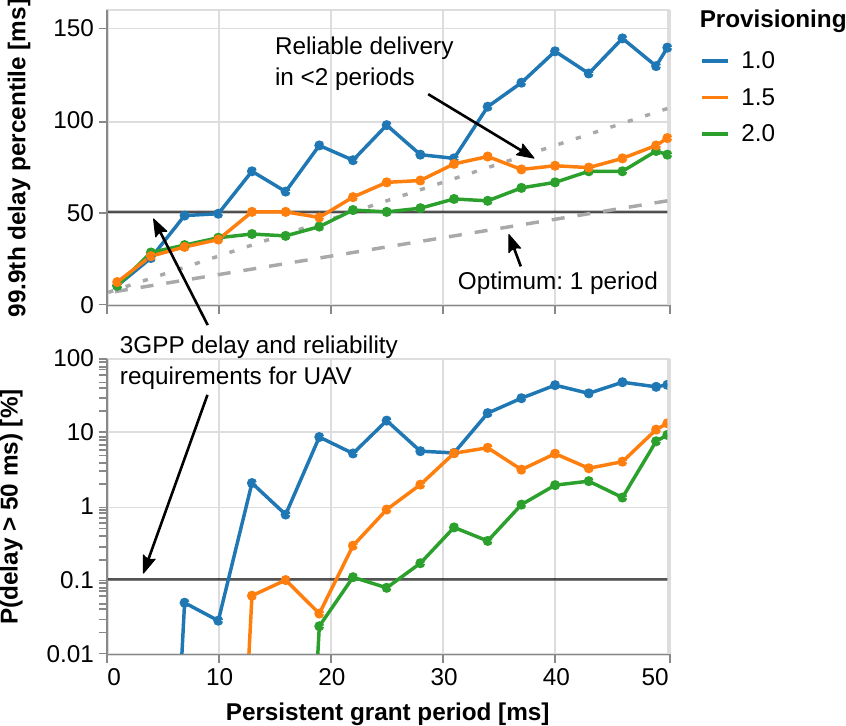}
    \caption{Reliability of the scheduling approach evaluated through the 99.9th percentile of delays and the relative frequency of delays exceeding 50~ms. The dashed line indicates one period, and the dotted line indicates two periods, offset by the \SI{6}{ms} baseline delay. 3GPP requirements can be met in a range of configurations, with grant periods up to 25~ms.}
    \label{fig:delay_reliability}
    \vspace{-4mm}
\end{figure}

The tails of the delay distribution behave similarly to its mean. Figure~\ref{fig:delay_reliability} shows the upper bound of delays with a relative frequency of \SI{99.9}{\percent} (top), and the probability of a message being delayed over \SI{50}{\milli\second} (bottom), in comparison to the requirements listed in~\cite{3gpp-ts36.777r15}. %
The lowest 99.9th delay percentile of \SI{10}{\milli\second} is achieved with the shortest scheduling period.
Upward of \SI{35}{\milli\second} period length, an overprovisioned allocation can deliver messages reliably in less than two periods, whereas no overprovisioning leads to approximately three periods.
The requirement of \SI{50}{\milli\second} latency (\SI{99.9}{\percent} probability) can be fulfilled with all factors, yet the longest permissible period increases with overprovisioning.

\begin{figure*}[t]
    \centering
    \vspace{2mm}
    \includegraphics{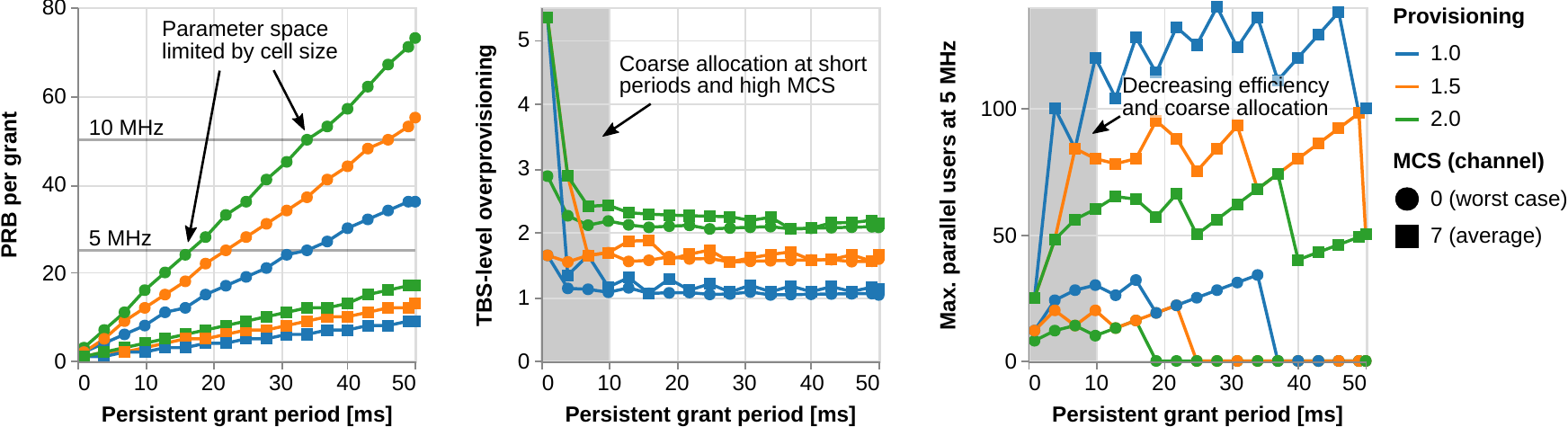}
    \caption{Resource metrics for the evaluated combinations of grant period and provisioning factor.}
    \label{fig:scalability}
    \vspace{-4mm}
\end{figure*}

\subsection{Scalability and Trade-Offs}
As shown in Figures~\ref{fig:delay_stats} and~\ref{fig:delay_reliability}, there is an overall expected trade-off between latency, reliability and capacity.
Figure~\ref{fig:scalability} shows resource metrics over the range of evaluated parameters both for the used \ac{MCS}~7 and for the most robust \ac{MCS}~0 required at the edge of the base station's range.
Note that short grant periods restrict allocation granularity, as depicted in Fig.~\ref{fig:scalability} (middle).
At the extreme of \SI{1}{\milli\second} periods, all parameterizations result in 1~block allocations, providing over five times as much resources as needed regardless of the provisioning factor.
At longer grant periods, resource usage converges towards $\alpha r$ as finer-grained allocation becomes possible.
Both mean latency and reliability can be traded for capacity through overprovisioning.
However, the idealized mean delay of $\frac{p}{2}$ only holds at higher periods with double provisioning. %
Table~\ref{tab:delay_reliability} shows a set of parameters satisfying delays under \SI{50}{\milli\second} with \SI{99.9}{\percent} reliability.
The top section shows configurations with least resource usage for each provisioning factor, while the bottom section lists the configuration with minimum mean delays, i.e., grants in each subframe. Note that the provisioning factor has no effect for resource allocation in each subframe, as all factors lead to effective over-allocation of resource by a large margin (see Fig.~\ref{fig:scalability}).
Among the resource-efficient parameterizations, 1.5 fold overprovisioning achieves the best delay statistics, indicating a local optimum among the configurations fulfilling the 3GPP requirements.
Note that a reduction in resources per user does not always entail an increase in total users.
The result of equation~(\ref{eq:grant_scalability}) over all parameterizations is shown in the right column of Fig.~\ref{fig:scalability}.
While the number of possible users increases with period length, larger grants allow fewer subscribers to be scheduled in a single subframe.

\begin{table}[b]
\vspace{-4mm}
\caption{Parameters satisfying 3GPP requirements with least resource usage and minimum mean delay}
\centering
\begin{tabular}{lcccr}
\toprule
& & \multicolumn{3}{c}{Delay (ms)} \\
\cmidrule(r){3-5}
Provisioning & Period (ms) & Mean & 99.9\% & Std. dev. \\
\midrule
1.0 & 10 & 15.9 & 49 & 7.6 \\
1.5 & 19 & 11.4 & 47 & 7.6 \\
2.0 & 25 & 18.6 & 50 & 8.8 \\
\midrule 
1.0, 1.5 and 2.0 & 1 & 6.1 & 10 & 0.9 \\
\bottomrule
\end{tabular}
\label{tab:delay_reliability}
\end{table}

\addtolength{\textheight}{-0cm}

\section{CONCLUSION \& OUTLOOK}
\label{sec:conclusion}

We presented our novel approach for utilizing high capacity mobile cellular communication systems to enable low latency guarantees for mission-critical \ac{UAV} command and control.
We developed a scheduler using real hardware and an open-source \ac{LTE} stack, to conduct empirical evaluations with the help of \ac{SDR} modules.
Evaluations demonstrate that our approach can consistently fulfill mission-critical communications requirements in the context of \ac{SAR} scenarios, even in the presence of background traffic.
Furthermore, our resource-guaranteed channel access scheme provides low latency down to a value of \SI{6}{\milli\second}, with high reliability.
However, achieving this requires a two- or threefold increase in reserved capacity.
Yet, the loss of capacity could be considered a minor inconvenience to ensure safe operation in aerial vehicle missions.
Future work will focus on optimizing the uplink grant allocation algorithm, for which our results already present a starting point in the form of local optima in the latency, capacity, and reliability domains.
Finally, different model-based packet generators will be used to adapt scheduling parameters to a wide variety of situations and applications.

\balance

\section*{ACKNOWLEDGMENT}
\footnotesize{
Parts of this work were supported by the German Federal Ministry of Education and Research (BMBF) in the projects LARUS (13N14133) and A-DRZ (13N14857).
}

\bibliographystyle{./IEEEtran}

\end{document}